\documentclass[conference,letterpaper,10pt,times]{IEEEtran}
\usepackage[affil-it]{authblk}
\usepackage{amsmath}
\usepackage{amssymb}
\usepackage{array}
\usepackage{balance}
\usepackage{booktabs}
\usepackage[font={small}]{caption}
\usepackage{enumitem}
\usepackage[multiple]{footmisc}
\usepackage{graphicx}
\usepackage[misc]{ifsym}
\usepackage{makecell}
\usepackage{multirow}
\usepackage[font={small}]{subcaption}
\usepackage{times}
\usepackage{titlesec}

\makeatletter
\def\@citex[#1]#2{\leavevmode
\let\@citea\@empty
\@cite{\@for\@citeb:=#2\do
{\@citea\def\@citea{,\penalty\@m\ }%
\edef\@citeb{\expandafter\@firstofone\@citeb\@empty}%
\if@filesw\immediate\write\@auxout{\string\citation{\@citeb}}\fi
\@ifundefined{b@\@citeb}{\hbox{\reset@font\bfseries ?}%
\G@refundefinedtrue
\@latex@warning
{Citation `\@citeb' on page \thepage \space undefined}}%
{\@cite@ofmt{\csname b@\@citeb\endcsname}}}}{#1}}
\makeatother

\newcommand{\paragraphheader}[1]{\textbf{#1.}}

\newcommand{\circlenumber}[1]{\textcircled{\raisebox{-0.65pt}{\footnotesize{#1}}}}

\begin{document}

\title{A Mobile Payment Scheme Using Biometric Identification with Mutual Authentication}
\author{\textbf{Jack Sturgess and Ivan Martinovic}}
\affil{Department of Computer Science, University of Oxford, Oxford, UK \\\{firstname.lastname\}@cs.ox.ac.uk}
\maketitle

\begin{abstract}
Cashless payment systems offer many benefits over cash, but also have some drawbacks. Fake terminals, skimming, wireless connectivity, and relay attacks are persistent problems. Attempts to overcome one problem often lead to another---for example, some systems use QR codes to avoid skimming and connexion issues, but QR codes can be stolen at distance and relayed. In this paper, we propose a novel mobile payment scheme based on biometric identification that provides mutual authentication to protect the user from rogue terminals. Our scheme imposes only minimal requirements on terminal hardware, does not depend on wireless connectivity between the user and the verifier during the authentication phase, and does not require the user to trust the terminal until it has authenticated itself to the user. We show that our scheme is resistant against phishing, replay, relay, and presentation attacks.
\end{abstract}

\begin{IEEEkeywords}
mobile payments, mutual authentication, visual channel, biometrics, identification, authentication
\end{IEEEkeywords}

\section{Introduction}\label{sec:Introduction}

Mobile payment systems have evolved rapidly over recent years, facilitated by advances in technology and driven by enhanced security and usability features~\cite{Huh2017}. One early barrier to adoption was the need for specialised hardware, where exclusionary business practices mandated the use of a dedicated point-of-sale terminal for each payment system. Merchants struggled to support them all, so brand loyalty and local trends led consumer decisions. This was resolved by the standardisation of NFC-enabled payments (near-field communication; for short-ranged connectivity); now, a single terminal can accept payments made using any NFC-enabled device---whether it be a payment card, a hand-held device, or a wearable device---across various payment systems.

The size, shape, and capabilities of point-of-sale terminals have also changed, and are continuing to change. Early terminals had a slit where magnetic strip payment cards could be swiped to be read; these were replaced with slots where Chip and PIN payment cards could be inserted to have their chip read and unlocked by the PIN (personal identification number; a short numerical password). Terminals typically have a screen where the payment amount is displayed, a keypad or touchscreen where the PIN can be entered, and, for NFC-enabled terminals, a flat surface where the payment device is to be tapped to communicate with the NFC module. They need a trusted execution environment where local cryptographic materials can be handled and stored securely and a means to connect securely to a back-end payment server. Many modern consumer devices satisfy these requirements and so can be used as terminals; new tap-on-phone applications enable smartphones and tablet computers to accept NFC payments, increasing the number and diversity of potential terminals and allowing the payment process to be integrated alongside other vendor services in all-in-one applications---\textit{e.g.}, some restaurants use table management systems installed on tablets that enable waiting staff to take orders, transfer orders to the kitchen, and handle payments in the same application.

The physical presence of a terminal can give a false perception of trust to users that the terminal is legitimate, but a rogue terminal or application can easily be dressed to look genuine in order to execute a MITM (man-in-the-middle) attack on unsuspecting victims. An effective countermeasure to MITM attacks is mutual authentication, where the payment system must authenticate itself to the user (inasmuch as it must prove that it is a secure communication interface to a trusted server) before the user is asked to enter any secret information. This principle has been deployed in some smartphone applications~\cite{Marforio2016} and online banking interfaces~\cite{Tangerine2017}, where a personalised greeting message is shown to the user before the system requires the password to be entered. However, we are yet to see such a feature in point-of-sale terminals in the wild.

It is difficult to prevent skimming and eavesdropping over wireless channels. The use of NFC can allow payment transactions to be initiated and information stolen without the user knowing~\cite{Lishoy2010,Murdoch2010} and attempts to limit its range to reduce the risk have been shown not to be reliable~\cite{Diakos2013,Kortvedt2009}. Some payment systems communicate over a visual channel between the user's device and the terminal, giving the user more control over the exchange and making it difficult for an attacker to intercept without being noticed. However, these systems typically encode the information into barcodes or QR (quick response) codes that can be read and stolen at long range using a camera with sufficiently high resolution and then used in a relay attack. Furthermore, even though a wireless connexion is not required during the authentication phase, the user's device must frequently connect to the server at other times to request new tokens since the communication with the terminal over a visual channel is unidirectional.

The use of biometrics is starting to replace the PIN to authenticate the user because it requires less effort from the user---in some cases directly, such as fingerprint-enabled payment cards, and in other cases indirectly, where payment cards are provisioned to a virtual wallet on a smartphone and the user must authenticate to the device using whatever capabilities it offers to gain access to the virtual wallet. Some systems~\cite{Lee2017,Mastercard2022} are trying to phase out cards and devices altogether and instead use face recognition to identify the user among registered users and then automatically bill the account associated with the matched user. However, an unregistered attacker could abuse this by getting matched to a random user and causing that user to get billed, so a smartphone application is required to verify the match, meaning that the user must carry a wirelessly-connected smartphone, which detracts from the potential usability gains that these systems hoped to offer.

In this paper, we encapsulate these common problems into a set of three system requirements and then we propose a novel payment scheme that satisfies them. Our scheme leverages biometric identification to maximise convenience for the user and provides mutual authentication to protect the user from rogue terminals. We show that our scheme meets these requirements and that it is resistant against phishing, replay, relay, and presentation attacks.

\section{System Design}\label{sec:visauth:systemdesign}

\subsection{System Requirements}\label{sec:visauth:systemrequirements}

The purpose of a payment scheme is to authenticate a user to a verifier via a point-of-sale terminal to authorise a payment. In addition to this, we want our scheme to overcome the problems that are commonly found in existing mobile payment systems. To ensure that we meet this objective, we derive the following three system requirements from these problems that our scheme must satisfy:
\begin{itemize}
	\item \textit{No specialised hardware}: the system may only impose hardware requirements on the terminal that can be satisfied by typical smartphones and tablet computers, so that the system is simple for merchants to deploy.
	\item \textit{No user-to-verifier connexion}: the system must not require any device of the user to communicate directly with the verifier during the authentication phase, since network connectivity cannot always be guaranteed.
	\item \textit{No expectation of trust}: the user must not be expected to trust the terminal, and therefore must not be expected to reveal any secret information to it, until the terminal has first authenticated itself (\textit{i.e.}, proved that it is legitimate) to the user.
\end{itemize}

\begin{figure}[!t]
	\centering
  	\includegraphics[width=0.9\linewidth]{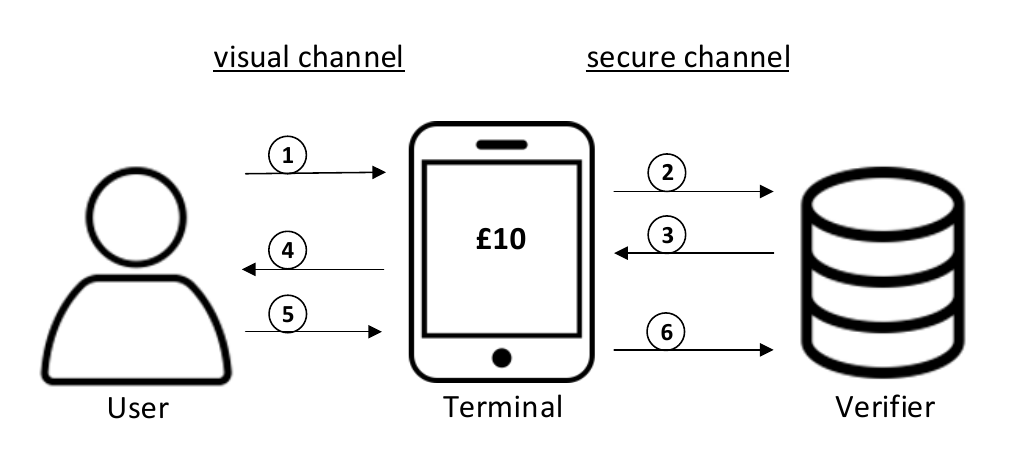}
   	\caption[The system model of our scheme during the authentication phase.]{The system model of our scheme during the authentication phase. The user presents his biometric trait(s) to the terminal~\circlenumber{1}, which extracts a feature vector and sends it to the verifier~\circlenumber{2}, which attempts to identify the user. The verifier returns the verification message associated with the account of the nearest matching user to the terminal~\circlenumber{3}, which displays it to the user to authenticate the terminal to the user~\circlenumber{4}. The user then enters his PIN to the terminal~\circlenumber{5}, which sends it to the verifier, which verifies the match to authenticate the user and then authorises the payment~\circlenumber{6}.}
   	\label{fig:visauth:systemoverview}
\end{figure}

\subsection{System Model}\label{sec:visauth:systemmodel}

We consider a system model in which a user is making a payment at a point-of-sale terminal in a typical setting (\textit{e.g.},~in a shop). The system consists of three components: a \textit{user} (the prover), a point-of-sale \textit{terminal}, and a \textit{verifier} (the back-end payment server that authorises the payment). The terminal is a commercial off-the-shelf device that the merchant is using to take the payment; it has a camera, a screen, an input mechanism (\textit{e.g.},~a keypad or a touchscreen), and an installed application that implements our scheme. The verifier is an authentication server maintained by the payment provider and is assumed to be trusted and secure. We assume that the terminal is registered to the verifier and that these devices have shared keys and established a secure channel over which to communicate. We assume that all cryptographic materials are stored securely on their respective devices.

During the enrolment phase, we assume that the user has a user device, such as a smartphone, that he uses to enrol into the system and to administer his account. We assume that there is a secure channel between the user device and the verifier over which they exchange materials. We assume that the user is able to access a trusted terminal, authenticated using the aforementioned secure channel, to submit a number of biometric samples from which his initial user template can be constructed. We assume that the biometric being used is face geometry, but other traits could be implemented with due consideration. The user adds a payment method to his account, chooses a PIN, and sets a verification message.

\begin{table*}[!t]
	\centering
	\begin{tabular}{ccccl}
		\toprule
		 & \textbf{} & \textbf{} & \textbf{} & \\
		 & \textbf{Known to or} & \textbf{Stored on} & \textbf{Stored on} & \\
		\textbf{Material} & \textbf{Inherent to User} & \textbf{Terminal} & \textbf{Verifier} & \textbf{Purpose} \\
		\midrule
		$a$ & (\checkmark) & (\checkmark) & $\times$ & payment amount; known at start of session \\
		$b$ & \checkmark & $\times$ & (\checkmark) & biometric feature vector; identifies the user to the verifier by nearest match \\
		$m$ & \checkmark & $\times$ & \checkmark & short alphanumeric string; authenticates the terminal to the user \\
		$PIN$ & \checkmark & $\times$ & \checkmark & short numerical password; authenticates the user to the verifier \\
		$k$ & $\times$ & \checkmark & \checkmark & secret key; secures communication between the terminal and the verifier \\
		\bottomrule
	\end{tabular}
	\caption[Summary of the cryptographic materials used in our scheme.]{Summary of the cryptographic materials used in our scheme.}
	\label{tab:summaryofmaterials}
\end{table*}

During the authentication phase, we assume that there is a visual channel between the user and the terminal, and a secure channel between the terminal and the verifier. The user interacts only with the terminal. The user initiates the protocol by presenting his biometric trait to the terminal. We assume that the integrity of the biometric sample is protected with appropriate liveness detection, as is standard practise in biometric-based systems. The terminal extracts a feature vector from the biometric sample and sends it to the verifier for identification. When the verifier identifies the user, it retrieves the verification message associated with his account and returns it to the terminal, which displays it to the user. The user verifies the message and enters his PIN to the terminal, which passes it to the verifier. The verifier checks the PIN to verify that the identification was correct and to authenticate the user. The verifier can then authorise the payment using the payment method associated with the account. We assume that the classifier used by the verifier has a low misclassification rate and that, when a user is correctly identified and authenticated, his template can be safely updated using the latest feature vector to counter the effects of drift (where a biometric trait changes over time, such as due to ageing). We assume that the verifier will reject simultaneous authentication sessions that are identified to be of the same user to prevent crossover. A visualisation of the authentication phase of our scheme is shown in Figure~\ref{fig:visauth:systemoverview}.

\subsection{Threat Model}\label{sec:visauth:threatmodel}

We consider an adversary that is attempting to make a payment at the expense of a legitimate user. We assume that the adversary can observe everything that is shared across the visual channel. We assume that the adversary can deploy rogue terminals and that these are dressed to look genuine. Our goal here is to authenticate the legitimate user without leaking any secret information, to facilitate legitimate mobile payment transactions, and to reject the adversary. We consider the following six types of attack:
\begin{itemize}
	\item \textit{Phishing attack}: the adversary has deployed a rogue terminal to trick a legitimate user into revealing his PIN.
	\item \textit{Replay attack}: the adversary is attempting to make a new payment by re-using (eavesdropped) messages that were previously sent between a legitimate user and the verifier.
	\item \textit{Relay attack (in-store)}: a legitimate user is attempting to make a payment at a rogue terminal that is passing his biometric trait (\textit{e.g.},~a captured image of his face) to an adversary who is attempting to use it to authorise a different payment at a legitimate terminal.
	\item \textit{Relay attack (skimming)}: while a legitimate user is \textit{not} involved in a transaction (\textit{e.g.},~he might be commuting on public transport or walking on a busy street), the proximate adversary is attempting to capture his biometric trait using a concealed rogue terminal so as to pass it to a distant accomplice who is attempting to use it to authorise a payment at a legitimate terminal.
	\item \textit{Presentation attack (particular victim)}: the adversary has observed the PIN and biometric trait of a legitimate user in a previous transaction and is attempting to make a payment by impersonating that user.
	\item \textit{Presentation attack (random victim)}: the adversary is attempting to make a payment as a random user.
\end{itemize}

In this work, we concentrate on how the proposed scheme can be used to defend against these attacks. We do not consider attacks that take place during the enrolment phase, attacks on the liveness detection system, attacks on the verifier, malware, or denial of service attacks.

\section{System Architecture}\label{sec:visauth:systemarchitecture}

\subsection{Cryptographic Materials}\label{sec:visauth:cryptographicmaterials}

During the enrolment phase, the user account is created and the user exchanges some materials with the verifier that are stored securely and later used in the authentication phase. A summary of these materials is shown in Table~\ref{tab:summaryofmaterials}.

$b$ is the biometric feature vector. During the enrolment phase, the user submits a number of biometric samples to a terminal, which extracts feature vectors and sends them to the verifier, which constructs a biometric template for the user. This template is stored as part of the user's account. During the authentication phase, the user provides a biometric sample to the terminal. The terminal extracts a feature vector $b$ and sends it to the verifier to identify the user. It is best practice for biometric data to be processed locally on the device that collects it and for only the feature vector to be transferred, due to the irrevocable nature of biometric data and the impact that theft may have on the security and privacy of the user (across this system and other systems).

$m$ is the verification message. During the enrolment phase, the user chooses a recognisable string and submits it to the verifier. This message is stored as part of the user's account. During the authentication phase, the verifier sends $m$ to the terminal so that the terminal can authenticate to the user before the user is asked to reveal any secret information. Since the adversary can see $m$ when it is displayed, it must be changed after each use for mutual authentication to hold. We assume that this is achieved by using seeded random string generators to generate the same messages at regular intervals (\textit{e.g.}, every minute) on both the user device and the verifier to remove the need for any user-to-verifier connexion---however, this requires the user to carry the user device. We consider alternative implementations that prioritise usability in Section~\ref{sec:discussion}.

$PIN$ is the PIN. During the enrolment phase, the user chooses a short (\textit{e.g.}, 4 digits), memorable PIN and submits it to the verifier. This PIN is stored as part of the user's account. During the authentication phase, the user inputs $PIN$ to authenticate to the verifier. More specifically, when the verifier has selected the candidate user that most closely matches $b$, the $PIN$ input by the user is used to verify the match. We assume that the user will enter $PIN$ on the terminal over a physical channel. Alternative implementations might explore the use of other channels to verify the match, such as having the user speak $PIN$ over an audio channel, gesticulate $PIN$ over a visual channel, or provide some input in response to a challenge on a user device. In an ideal implementation of the scheme, only \textit{salted hashes} of $PIN$ should be stored, transferred, and compared in order to mitigate any damage from attacks against the verifier.

$k$ is a secret key shared between the terminal and the verifier. We assume that this is exchanged as part of the secure channel and is cryptographically secure.

\begin{figure*}[!t]
	\centering
	\begin{tabular}{c}
		\subfloat{
			\hspace{-2em}
			\includegraphics[width=0.94\linewidth]{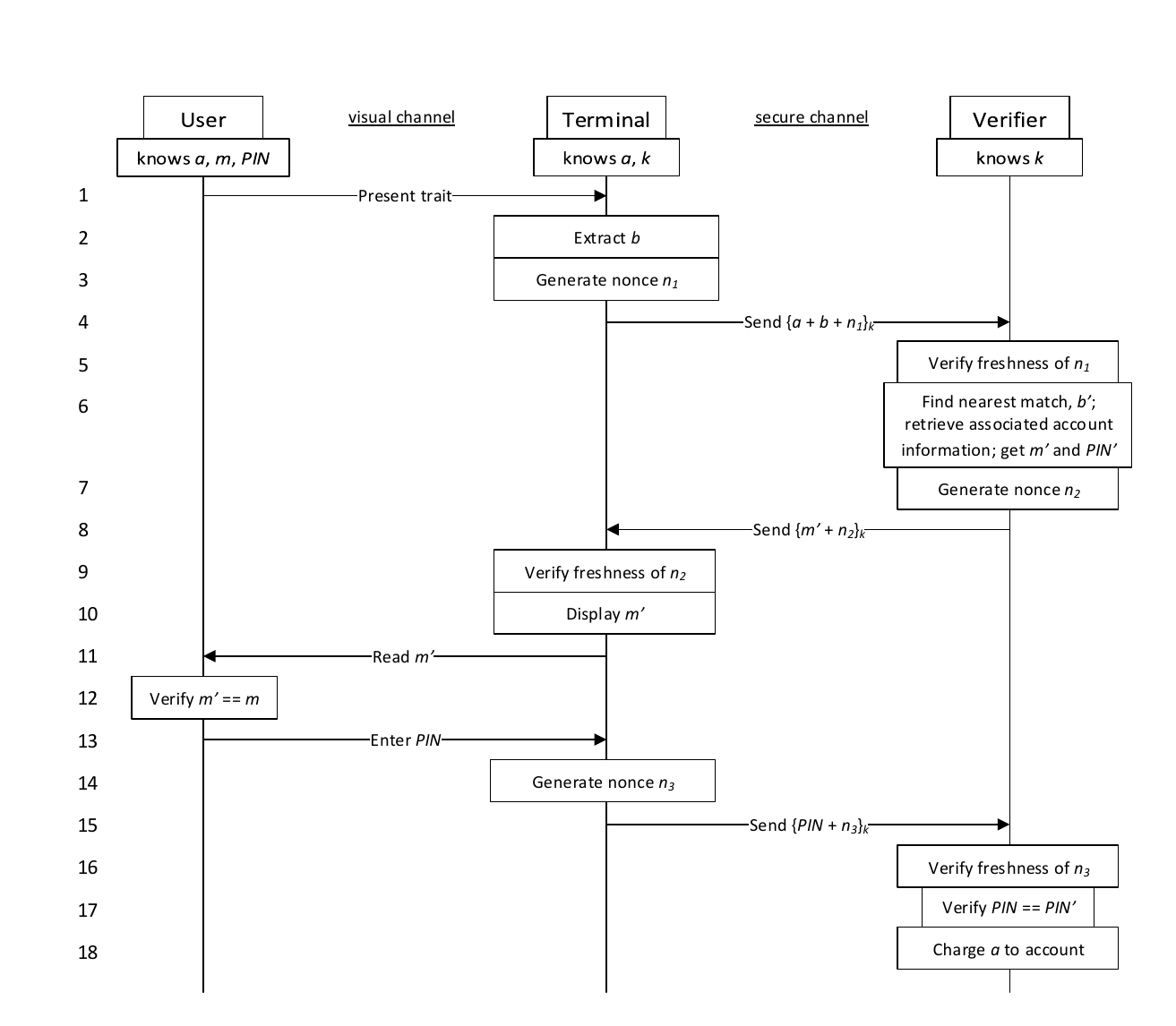}
		}
   	\end{tabular}
   	\caption[The authentication protocol of our scheme.]{The authentication protocol of our scheme.}
   	\label{fig:architectureauthentication}
\end{figure*}

\subsection{Biometric-based Identification}\label{sec:biometricbasedidentification}

We exploit the property that biometrics can be classified in a one-to-many manner (\textit{i.e.},~for identification purposes) to enable the user to bypass initially having to reveal any concrete information to the terminal. The user presents only his biometric trait to the terminal---in our case, this is his face, which is freely observable in public. The user's account information, such as his account number, is known only to the user and the verifier and does not need to be transferred during the authentication phase.

Biometric classification can result in false positives. If the identification task performed by the verifier returns a false positive, this will be caught when the user is presented with a message that does not match $m$. If the user tries to proceed anyway, he will fail because he does not know the PIN associated with the mismatched account. For usability, one approach to resolving this problem could be for the classifier to return a shortlist of candidate users ordered by how closely they match $b$ and for the subsequent message verification steps to repeat, iterating through the list, until the correct $m$ is displayed and verified. However, this approach would enable an attacker to collect verification messages of near-matching users that could be used in a phishing attack. Moreover, it would train users to tolerate false matches. In either case, the mutual authentication property would be undermined. For security, the protocol should instead terminate if the verification message is not as expected. The user can then restart it by presenting to the terminal again to give the system a fresh opportunity to identify him correctly.

\subsection{Authentication Protocol}\label{sec:authenticationprotocol}

During the authentication phase, a visual channel between the user and the terminal and a secure channel between the terminal and the verifier are required for the system to achieve mutual authentication. Figure~\ref{fig:architectureauthentication} shows the authentication protocol and the following steps describe it.

\paragraphheader{Steps~1 to~2: User Presents Trait to Terminal} The user approaches the terminal to initiate a payment transaction in the amount of $a$. The user presents his biometric trait to the terminal. The terminal samples the trait and extracts a feature vector, $b$.

\paragraphheader{Steps~3 to~12: Terminal Authenticates to User} The terminal sends $a$ and $b$ to the verifier over a secure channel. Nonces are used so that the freshness of messages can be verified by the receiver; we assume that these will take the form of timestamps. The verifier identifies the user by performing a one-to-many lookup of $b$ amongst the user templates of all users registered on the system. When the nearest match, $b'$, is identified, the verifier retrieves the account information of $b'$ and gets its verification message, $m'$, and PIN, $PIN'$. The verifier sends $m'$ to the terminal, which displays it to the user. The user verifies $m'$ by performing a string comparison against the expected message, $m$. This demonstrates to the user that the terminal is securely communicating with and trusted by the verifier and so authenticates the terminal to the user before the user is required to reveal any secret information.

\paragraphheader{Steps~13 to~18: User Authenticates to Verifier via Terminal} The user enters $PIN$ on the terminal, which sends it to the verifier. The verifier verifies $PIN$ against the expected $PIN'$. This authenticates the user as the identified user to the verifier and enables the verifier to process the payment of $a$ using the payment method associated with the identified user account.

\section{Security Analysis}\label{sec:securityanalysis}

Our scheme meets all of the system requirements. It requires no specialised hardware: the terminal needs only a camera, a screen, and a touchscreen or keypad for input; these requirements can be satisfied by any modern smartphone or tablet, making it easily deployable. The user is identified and authenticated to the verifier via the terminal, using a visual channel, without requiring a direct connexion between the user (or any user device) and the verifier. Finally, the authentication protocol ensures that the terminal demonstrates to the user that it is connected to and trusted by the verifier before the user is required to reveal $PIN$.

\paragraphheader{Phishing Attack} For the phishing attack, the adversary attempts to have a legitimate user reveal $PIN$ by deploying a rogue terminal. At Step~10, the terminal must display $m$ to the user before the user reveals $PIN$. As long as $m$ is changed after each use, the rogue terminal will not be able to achieve this. Therefore, our scheme provides resistance against phishing attacks.

\paragraphheader{Replay Attack} For the replay attack, the adversary attempts to authorise a repeat payment by re-sending encrypted messages sent between the terminal and the verifier during a previous transaction. Nonces are included in every encrypted message sent over the secure channel to enable their freshness to be checked, so the attack will fail at Step~5. Therefore, our scheme provides resistance against replay attacks.

\paragraphheader{Relay Attack} For the in-store relay attack, the adversary attempts to authorise a different payment by passing a copy of the legitimate user's biometric trait. For the skimming relay attack, the adversary attempts to have an accomplice authorise a payment by capturing the biometric trait of an oblivious legitimate user. In each case, owing to the use of a visual channel, the legitimate terminal being used by the accomplice is able to validate the authenticity of the user. The copied biometric will fail the liveness check and be rejected, so the attack will fail at Step~2. Therefore, our scheme provides resistance against relay attacks.

\paragraphheader{Presentation Attack} For the presentation attack on a particular victim, the adversary knows the victim's PIN and attempts to mimic his biometric trait so as to impersonate him to a legitimate terminal. Biometric identification is weaker than authentication inasmuch as the attacker only needs to achieve being matched nearer to the intended victim than to some other user. We can minimise this discrepancy by tightening the decision threshold of the classifier to ensure that the matching must meet a certain minimum accuracy, akin to authentication. The tighter this is set, the more it will increase the FRR (\textit{i.e.},~the greater the gains in security, the greater the cost to usability). This will increase the effort required of the adversary. An implementation of the scheme can further increase the effort required of the adversary by using multiple biometric traits, since he would need to achieve being the nearest match for all of the traits simultaneously.

For the presentation attack on a random victim, the adversary simply presents his own biometric trait, either modified or not, so that the system matches him to a random victim in Step~6. We assume that the adversary is not registered to the system and that a match is found, even with a tightened decision threshold. The adversary does not know the PIN of the random victim, so the attack will fail at Step~17. The adversary may attempt to perform a brute force guessing attack against the PIN by presenting in the same manner repeatedly to generate the same victim each time. An ideal implementation of the scheme should use common anomaly detection and throttling techniques to defend against guessing attacks.

The first of these attacks does not work at scale, because the adversary must expend effort to obtain the victim's PIN, and is defeated by further increasing the effort (cost) required. The second does work at scale, but is defeated by the PIN. In each case, our scheme provides resistance against presentation attacks.

\section{Discussion}\label{sec:discussion}

\paragraphheader{Convenience} Our scheme enables the user to make a payment without needing to carry any form of cash, payment card, or user device (although he may need to consult his user device to verify $m$, depending on how it has been implemented). Furthermore, since the user is identified as part of the process, any relevant status checks can be made automatically against the information held on record. This means that the user does not need to carry loyalty cards, discount coupons, or proof of age or membership---as these can all be applied upon identification.

For security, we have assumed that seeded random string generators on the user device and the verifier generate a fresh $m$ every minute. To improve usability, an implementation of the scheme might consider ways to increase the size of the interval between changes of $m$ to free the user from the requirement that he must carry a user device during the authentication phase. This could include the use of message templates, along with shapes or colours to increase the message space, so that what needs to be memorised is more user-friendly, rather than a random string. For example, the system might allow the user to create or select a rule pertaining to the structure of an expected message that is valid for a day, then the verifier would randomly generate a fresh string every transaction that satisfies the rule so that the user only needs to verify that it fits the template (\textit{e.g.}, `a valid 5-letter word followed by a green triangle'). The user would then memorise the rule before a shopping session and not need to further consult his user device. The adversary could replay such a message to perform a phishing attack, but not at scale, so the gains in usability may be worth the risks to security.

\paragraphheader{Asymmetrical Channel} The visual channel enables asymmetric communication, as the capabilities required for sending information are different to those required for receiving it. Each party can either display to or read from the channel depending on its capabilities. This means that there are constraints on what each party can do to each other and parts of the system can be restricted to unidirectional communication. Our scheme leverages this property in Steps~1 and~2, where the user presents his biometric trait and the terminal can only read it, and in Steps~10 and~11, where the terminal displays $m'$ and the user can only choose whether or not to verify it.

\paragraphheader{Contextual Awareness} The capabilities that can be used to read information from the visual channel can also collect incidental information from the surrounding environment. Depending on its position, the camera on the terminal can capture additional information around the user that could be used to facilitate advanced fraud detection techniques, such as verifying that the terminal is operating in the expected environment. An implementation of the scheme might leverage this property by passing an image of the scene to the verifier; expected objects, markers, or lighting effects could be placed in the environment as a form of signature, or a clock could be placed in the environment such that the time captured in the image could be extracted and cross-checked with the timestamps used as nonces to strengthen the assertion of freshness with an independent factor. A sophisticated adversary could still fabricate the entire environment, but each step would increase the effort required of the adversary and present a potential point of failure for an attack.

\paragraphheader{Privacy Risk Mitigation} The use of a visual channel---especially when collecting peripheral information---poses a risk to the privacy of the user. Any images sent to the verifier should have their utility weighed against their potential impact on privacy. Countermeasures to mitigate privacy leakage from images include reducing the resolution and blurring unnecessary details before sending. To protect the biometric data of the user, biometric traits should only be processed locally on the terminal and should be obscured from any images sent to the verifier.

\section{Related Work}\label{sec:relatedwork}

\paragraphheader{Identification-based Systems} With regard to the use of biometric identification as part of an authentication system, some payment providers have trialled the technique with the promise of improved convenience for the user. Smile-to-Pay~\cite{Lee2017}, developed by Ant Financial for AliPay, uses a 3D camera to capture the user's facial likeness, perform liveness detection, and identify the user within 2~seconds. The system then sends a verification request to the user's smartphone that requires a timely response to verify the match. Biometric Checkout Program~\cite{Mastercard2022}, developed by Mastercard, operates in a similar manner, allowing the user to identify himself to the terminal over a visual channel using either his face or palm. Both of these systems require a user-to-verifier connexion to verify the match. To the best of our knowledge, we are the first to propose the use of biometric identification to facilitate mutual authentication and to do so without requiring a user-to-verifier connexion.

\paragraphheader{Visual Channel} With regard to the use of a visual channel, some existing mobile payment systems have explored the use of a QR code to pass information between a user device and the terminal. In Yoyo Wallet~\cite{Yoyo2017}, the user must first authenticate to a smartphone application using a PIN and can then access a QR code that contains tokenised payment information. To make a payment, the QR code is shown to the terminal over a visual channel and can be used up to three times before it expires. When the user's smartphone next connects to the Yoyo cloud server, where the user's virtual wallet is stored, a new QR code is downloaded. The limited number of uses per QR code mitigates the damage from theft, but makes this payment system more dependent than typical tap-and-pay systems on a user-to-verifier connexion between transactions. WeChat~\cite{WeChat2017} and AliPay~\cite{AliPay2017}, both widely used payment systems in China, support the use of QR codes and barcodes to transfer information. VisAuth~\cite{Sturgess2017} embeds information into an image as a robust watermark to send it over a visual channel. However, the system state on the user device can become desynchronised from that on the verifier---while the authors describe this as a benefit, since it unavoidably draws attention to an attack, it can also happen if the protocol is interrupted at various steps, providing plausible deniability to an attacker and making the wider scheme impractical. All of these systems transfer confidential information over a visual channel, whereas we transfer authentication information. Our scheme makes broader use of the visual channel by observing the user's biometric trait(s) to identify him to the verifier, where his payment information is stored, rather than encoding the payment information directly into a visual token.

Smart city transport networks, such as Oxford SmartZone~\cite{OxfordBus2022}, enable bus tickets to be purchased in advance and delivered as QR codes to the user's smartphone application. Instead of buying a ticket from the driver, the user presents the QR code to a terminal on the bus to expedite boarding. In this case, a product, rather than the user, is being authenticated, so the threat model primarily considers theft. The user can only be logged in to one device at a time to prevent account-sharing on multiple devices and the screen contains additional, animated elements that are verified by the terminal to prevent token-sharing using a screenshot.

\paragraphheader{Mutual Authentication} With regard to mutual authentication, related works in the field of mobile cloud computing have focused on the mutual authentication of the user device and the verifier in a general setting without consideration for any other components that might be involved in the system, such as the point-of-sale terminal in our case, that also need to be authenticated before the user should be expected to reveal any secret information. Dey \textit{et al.}~\cite{Dey2018} proposed a scheme that relies on the location of the user device and the current time at the verifier, and so requires a persistent user-to-verifier connexion. Other works~\cite{Munivel2019,Olakanmi2018,Purnomo2016} have proposed schemes in which mutual authentication is achieved via a trusted third party. These schemes require there to be a persistent connexion between the user device and the third party from the start of the transaction. We have sought to avoid this for the same reason that we avoid a user-to-verifier connexion.

\section{Conclusion}\label{sec:conclusion}

In this paper, we proposed and analysed a novel mobile payment scheme based on biometric identification that operates over a visual channel. We showed that our scheme (i)~requires no specialised hardware, imposing only minimal requirements on the terminal that can be satisfied by most commercially available smartphones and tablet computers, to ease deployment, (ii)~requires no user-to-verifier connexion during the authentication phase, such that it remains usable regardless of wireless connectivity, and (iii)~mutually authenticates the terminal and the verifier to the user before he is asked to reveal any secret information to authenticate himself. We explored the properties that a visual channel provides and we showed that our scheme is extensible in various ways depending on the needs of the wider system in which it is implemented. Furthermore, our scheme provides a number of conveniences to the user, such as not having to carry any payment or loyalty cards, and provides resistance against phishing, replay, relay, and presentation attacks.

\section*{Acknowledgement}

This work was supported by Mastercard and the Engineering and Physical Sciences Research Council [grant number EP/P00881X/1]. The authors would like to thank these organisations for their support.

\balance

\end{document}